# Apparent Ferromagnetism in Exfoliated Ultra-thin Pyrite Sheets

Anand B. Puthirath[1†*], Aravind Puthirath Balan[2†,] Eliezer F. Oliveira[1,3,4†], Vishnu Sreepal[2], Francisco C. Robles Hernandez[1,5], Guanhui Gao[1], Nithya Chakingal[1], Lucas M. Sassi[1], Prasankumar Thibeorchews [1], Gelu Costin[7], Robert Vajtai[1], Douglas S. Galvao[1,3,4] Rahul R. Nair[2] and Pulickel M. Ajayan[1*]

[1]Department of Material Science and NanoEngineering, Rice University, Houston, Texas, 77005, United States.

[2]Department of Chemical Engineering and Analytical Science (CEAS) & National Graphene Institute (NGI), University of Manchester, Manchester, M13 9PL, UK

[3] Group of Organic Solids and New Materials, Gleb Wataghin Institute of Physics, University of Campinas (UNICAMP), Campinas, SP 13083-970, Brazil

[4] Center for Computational Engineering & Sciences (CCES), University of Campinas (UNICAMP), Campinas, SP 13083-970, Brazil

[5]Mechanical Engineering Technology Program, Department of Engineering Technology, College of Technology, University of Houston, TX, 77204, USA

†These authors contributed equally

Email: ajayan@rice.edu, anandputhirath@rice.edu

**Abstract**

Experimental evidence for ferromagnetic ordering in isotropic atomically thin two-dimensional crystals has been missing until a bilayer $Cr_2Ge_2Te_6$, and a three-atom thick monolayer $CrI_3$ are shown to retain ferromagnetic ordering at finite temperatures. Here, we demonstrate successful isolation of a non-van der Waals type ultra-thin nanosheet of $FeS_2$ derived from naturally occurring pyrite mineral ($FeS_2$) by means of liquid-phase exfoliation. Structural characterizations imply that (111) oriented sheets are predominant and is supported theoretically by means of density functional theory surface energy calculations. Spin-polarized density theory calculations further predicted that (111) oriented three-atom thick pyrite sheet has a stable ferromagnetic ground state different from its diamagnetic bulk counterpart. This theoretical finding is evaluated experimentally employing low temperature superconducting quantum interference device measurements and observed an anomalous ferromagnetic kind of behavior.

**Key Words**: Non-van der Waals exfoliation, Pyrite, liquid exfoliation, anomalous magnetic ordering

**TOC Graphic**

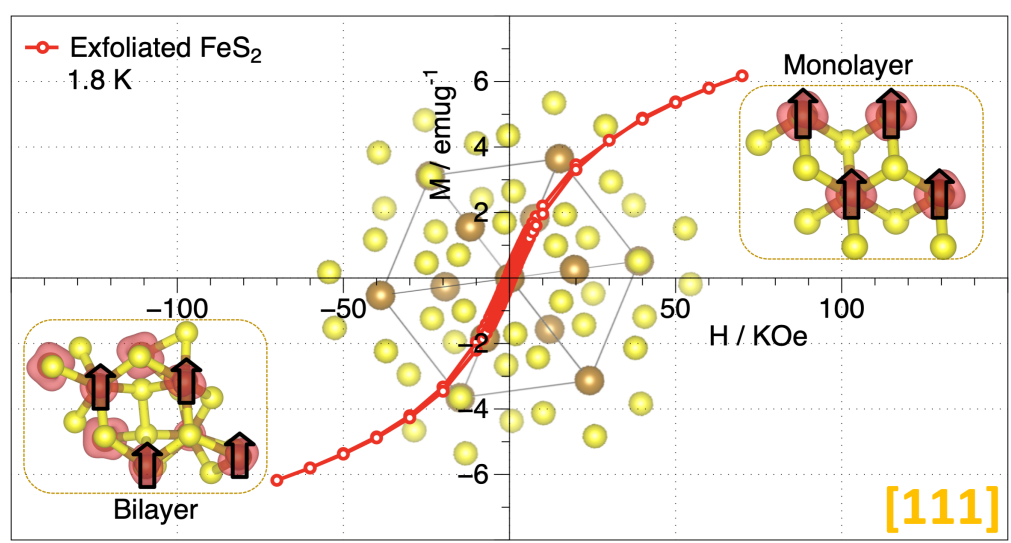

**Introduction**

Freestanding two-dimensional (2D) crystals provide unique opportunities for studying fundamental physics as well as for technological advancements.[1] The discovery of $Cr_2Ge_2Te_6$ and $CrI_3$, a couple of van der Waals (vdW) 2D materials that are capable of retaining ferromagnetic ordering at finite temperatures is groundbreaking and laid the foundation for wide research in 2D ferromagnetic materials and their heterostructures.[2-3] 2D magnetic crystals are susceptible to external stimuli such as mechanical stress, electric field, light incidence and chemical modifications which can be exploited effectively to engineer their magnetic properties for device applications.[4] In addition to that, ability to form heterostructures makes them excellent candidate materials to study proximity effects by stacking with other vdW 2D electronic, valleytronic and photonic crystals. Such heterostructures may result in new and novel phenomena such as multiferroicity, unconventional superconductivity and quantum anomalous Hall effect.[5-8] More vdW 2D atomic crystals emerged, namely $Fe_3GeTe_2$ and $Fe_5GeTe_2$, which are found to have ferromagnetic ordering close to room temperature. Ongoing investigations are aimed at discovering new and novel 2D atomic crystals capable of retaining ferromagnetic ordering at finite temperatures.

In addition to the exploitable robust properties, the prerequisite criteria of a potential 2D material for future applications are the abundance, the environmental friendliness and the cost-effectiveness of its bulk precursor. Naturally occurring minerals are the tangible choice of precursor materials that satisfies all the three aforementioned requirements. The prominent examples are vdW materials such as graphite and molybdenite from which graphene and 2D molybdenum disulfide ($MoS_2$) – two of the most explored materials in the entire 2D family were isolated.[9-10] Very recently, 2D materials derived from naturally occurring minerals are again coming back into the limelight in the form of franckeite.[11-13] Non-van der Waals (n-vdW) 2D materials are gaining popularity owing to their huge potential in catalytic applications and storage applications.[14-16] Recent investigations show that iron-based naturally occurring minerals are good precursor materials for envisaging new and novel n-vdW 2D materials for magnetic and catalytic applications.[17-19] Two-dimensional magneto-photoconductivity observed in non-van der Waals manganese selenide adds to the excitement.[20] Liquid-phase exfoliation in a suitable organic solvent is proved to be an effective method for exfoliation of these n-vdW mineral ores, which was already an established method for exfoliation of vdW 2D materials.[21]

Pyrite ($FeS_2$) is the abundant and the most common form of sulfides and is widely explored for photovoltaic applications.[22] A small bandgap energy (~0.95 eV) and reasonably good optical absorption coefficient (~ 6 x $10^5$ $cm^{-1}$) qualify pyrite to be used as an excellent candidate material to absorb a broad spectrum of solar light.[23] Besides photovoltaic applications, pyrite is also an excellent cathode material in lithium-ion batteries.[24] These remarkable properties along with cost-effectiveness, eco-friendliness and abundance make it a perfect material for nano-engineering with a view to enhance performance and check new phenomena, especially confinement effects in the 2D regime. Moreover, layered manganese-containing sulfides and selenides ($MnX_2$ ; X=S, Se) with the pyrite-type structure were already demonstrated to have high-temperature ferromagnetism down to monolayers, whereas their bulk counterparts are low-temperature antiferromagnetic materials.[25-26] Pristine $FeS_2$ is diamagnetic, and the prospect of electrically induced ferromagnetism in 'zero-spin' $FeS_2$ via ionic-liquid gating is one the recent advancements.[27] In this work, we successfully isolated ultrathin $FeS_2$ sheets from a non-layered naturally occurring mineral ore pyrite (Iron (II) sulfide) by means of liquid exfoliation. Our theoretical density functional theory (DFT) calculations suggest that (111) oriented three atom thick $FeS_2$ sheet is capable of retaining ferromagnetic ordering at finite low temperatures above 0K, whereas the bulk natural ore is diamagnetic. The theoretical findings are evaluated experimentally by low temperature magnetic measurements and the results are discussed.

**Materials and Methods**

**Exfoliation of $FeS_2$ nanosheets from bulk mineral ore pyrite:** Pyrite mineral (**Figure S1a**) ore was obtained from Department of Earth, Environmental and Planetary Sciences, Rice University and was ground to small grains of several microns. 20 mg of the powder was dispersed in 200 ml of N, N – dimethylformamide (DMF) solvent and subjected to ultrasonic bath cavitation for approximately 25 hours. A red-colored dispersion consisting of ultra-thin $FeS_2$ nanosheets was obtained as a result of extensive sonication, see **Figure S2**. Ultracentrifugation of the red-colored solution was performed at an RCF value of 10000 ($\times g$) for 15 minutes and the supernatant was filtered to obtain a pale red dispersion of ultrathin sheets of $FeS_2$. The solution thus obtained was vacuum filtered to obtain few milligrams of ultra-thin $FeS_2$ nanosheets which was used as such for further characterizations.

**Analytical Techniques:** The Electron Probe Micro-Analysis (EPMA) was performed at Rice University, Department of Earth, Environmental and Planetary Sciences Science using a Jeol JXA 8530F Hyperprobe, equipped with a field emission assisted thermo-ionic (Schottky) emitter and five Wavelength Dispersive Spectrometers (WDS). The analytical conditions used were 15 kV accelerating voltage, 20 nA beam current. Powder X-ray diffraction was obtained using a Rigaku Smart-lab X-ray Spectrometer. The Raman measurements were obtained employing Renshaw inVia Raman microscope with a 50x objective using He-Ne laser of wavelength 632.8 nm with a spot size of 1μ and the elemental analysis was performed in PHI Quantera XPS machine. AFM imaging was performed using a Bruker Dimension FastScan AFM operating in peak force tapping mode. The imaging was carried out using silicon nitride cantilevers, in air and at room temperature. The samples for the AFM images were prepared by drop-casting diluted dispersions onto $Si/SiO_2$ substrates. The low-resolution TEM and HRTEM images were obtained from an advanced Titan transmission electron microscope. For DC magnetic measurements, a magnetic property measurement system (MPMS 3) by Quantum Design was used.

**Theoretical tools:** The density functional theory (DFT) simulations were performed with a projector augmented wave potential (PAW) in a generalized gradient approximation (GGA) with the Perdew, Burke, and Ernzerhof (PBE) exchange-correlation functional.[28] A Monkhorst-Pack *k*-mesh of 4 x 4 x 1 was used to sample the Brillouin zone for geometry optimizations. To evaluate the electronic properties, we increase the Monkhorst-Pack *k*-mesh to 12 x 12 x 1. The Kohn−Sham orbitals were expanded in a plane-wave basis set with a kinetic energy cutoff of 70 Ry (~ 950 eV) and a Hubbard parameter[29-32] U = 2.0 eV was used for Fe atoms. This value was used in other 2D Fe structures (hematene).[17] The DFT simulations were performed with the computational code Quantum Espresso.[33] The exfoliation energies were estimated using the proposed method by J. H. Jung *et al.*[34]

**Results and Discussions**

Pyrite ($FeS_2$) has an isometric structure consisting of $Fe^{2+}$ ions surrounded by six octahedral coordinated sulfur ions. Pyrite is characterized by the presence of dianion $S_2^{2-}$ along with iron atoms occupy fcc sites with halite (NaCl) structure with $Pa\bar{3}$ space group.[35] From the surface energy calculations along with competing interlayer spacing and surface broken-bond density considerations, it is apparent that pyrite has three favorable cleavage/parting planes – (100), (110) and (111) oriented in [100], [110] and [111] directions respectively.[36] Exfoliation energy was theoretically calculated for pyrite along these energetically favorable planes using density functional theory (DFT) and are summarized in **Figure 1**. The DFT simulations obtain values of 0.48 eV/atom, 0.19 eV/atom, and 0.50 eV/atom, for (100), (111), and (110) planes (single layered) respectively suggesting that the (111) is the easier to exfoliate. A (111) faceted 2D $FeS_2$ sheet consists of a layer of Fe atoms sandwiched between two layers of sulfur atoms.

These observations motivated us to examine the experimental validity of the findings and are described in the following section.

The bulk precursor mineral pyrite is an electron probe micro-analysis (EPMA) natural standard and is compositionally homogeneous and inclusion-free except a few impurity atoms at ppm level (EPMA analysis, **Table S1**). Powder X-ray diffraction (XRD) was carried out on ground pyrite ore to confirm the crystallinity and purity and is shown in Figure 2A (top). The peaks are well matching with the $FeS_2$ pyrite phase (JCPDS No. 65-3321).[37] Later a few milligrams of exfoliated $FeS_2$ sheets collected via ultracentrifugation followed by vacuum filtration of the supernatant was also subjected to powder XRD analysis. The acquired result is depicted in **Figure 2a** (bottom), which is again comparable to bulk crystal (JCPDS No. 65-3321) but with broad peaks and different intensity ratios for various crystal orientations.

The Phase purity of both bulk and nanosheets was confirmed by Raman Spectroscopy (**Figure 2b**) with peaks centered at 340 and 375 cm$^{-1}$ corresponding to $E_g$ and $A_g$ symmetry modes of pyrite.[38] Peaks corresponding to other iron sulfides phases such as troilite, marcasite, and pyrrhotite are absent. Raman signature of exfoliation is evident in the intensity ratio of $A_g$ and $E_g$ modes ($I_{Ag}/I_{Eg}$). For bulk pyrite, $I_{Ag}/I_{Eg}$ is greater than unity whereas the nanosheets have an $I_{Ag}/I_{Eg}$ ratio less than unity. This is straightforward to understand since $A_g$ modes are due to out of plane vibrations, whereas $E_g$ modes are from in-plane vibrations. In n-vdW 2D crystals, the in-plane vibrations are predominant as compared to out of plane modes.[17] Another important feature in the Raman spectrum of $FeS_2$ nanosheets is the presence of low frequency (< 200 cm$^{-1}$) shear modes which is a clear indication of interlayer coupling similar to vdW 2D crystals, and is absent in the bulk 3D pyrite crystal.[39] X-ray photoelectron spectroscopy (XPS) measurements are carried out in order to elucidate the oxidation state of Fe and S and their atomic ratio (survey spectra, **Figure S3**) and are given in **Figure 2c-d**. Iron and sulfur peaks in exfoliated $FeS_2$ correspond to binding energies expected for pyrite; Fe2p3/2 at 707.0 and S2p3/2 and S 2p1/2 peaks at 162.6 and 163.9 eV respectively whereas a low-intensity Fe2p3/2 about 709.0 is due to surface states.[40-41] The exfoliated sheets were transferred onto a Si/SiO$_2$ substrate by drop-casting, and the thickness of individual sheets was measured using an atomic force microscope (AFM) and one of the images that shows the efficacy of exfoliation in obtaining the ultra-thin sheets is given in **Figure 2e**. It is evident from the AFM micrograph that the sheets are ultra-thin, and a stack of thick sheet (few layers, **Figure 2f**) on a tri-layer (**Figure 2g**) could be clearly distinguished from the corresponding line scans. Also, the thickness of several exfoliated sheets was measured, and are appended in the **Figure S4.**

The exfoliated FeS$_2$ sheets from the supernatant after ultracentrifugation were drop casted onto a holey carbon grid and were subjected to transmission electron microscopy (TEM) analysis. Low and high magnification transmission electron microscope (TEM) images, high-resolution TEM (HRTEM), fast Fourier transform (FFT), inverse fast Fourier transform (IFFT) and energy dispersive spectroscopy (EDS) images are presented in **Figure 3.** The low magnification (**Figure 3 a, b**) and high magnification (**Figure 3 c & d**) TEM images signify well exfoliated nanosheets and their contrast confirms the ultra-thin nature (**Figure 3a**) and its high homogeneity is evident in the high angle annular dark field (HAADF) image (**Figure 3b**). The higher magnification image in (**Figure 3d**) shows highly ordered regions along with their respective FFT in the inset, demonstrating that the arrangement of atoms in this selected area is along the (111) plane. The IFFT of that selected area yields an HRTEM (**Figure 3e**) that confirms the (111) preferential orientation of the exfoliated nanosheets. The TEM analysis reinforces the finding through the exfoliation energy calculations employing DFT that the exfoliation energy is least for (111) oriented planes. The modelled (111) projection plane using Vesta® software was made using the crystallographic information file (CIF) file for pyrite with the $Pa\bar{3}$ [205]space group space group having cubic crystal structure. The Vesta® projection is superimposed on the atomic resolution micrograph. The orientation of the zone axis direction is [111]. TEM EDAX (**Figure 3g-i**) confirms the purity of the nanosheets and is further supported by EPMA analysis given in the supporting information (**Table S2**). Extended TEM images and analysis are also provided in the **Figure S5.**

A monolayer pyrite, isostructural to 2D ferromagnetic VSe$_2$,[42] was claimed to be a potential material capable of exhibiting magnetic ordering at finite temperatures.[29] To elucidate the magnetic behavior of a 2D FeS$_2$ system, we performed spin-polarized simulations in a density functional theory (DFT) level,[28, 30-33] considering a layer of 2D FeS$_2$ oriented in [100], [111], and [110] directions, with initial unit cells containing 6 (2 FeS$_2$ formulas), 12 (4 FeS$_2$ formulas), and 12 (4 FeS$_2$ formulas) atoms, respectively. The complete set of data obtained for each 2D FeS$_2$ optimized in different magnetic states are presented in the Supplementary Information (see the *Supplementary Information* for the initial unit cells together with their initial lattice parameters **Table S3**). As we are considering 2D layers, the **c** unit cell axis was chosen to be 14.00 Å for all systems in order to avoid spurious interactions between neighboring structures due to the periodic boundary conditions. We performed the geometry optimizations for each system considering three different magnetic states: diamagnetic, ferromagnetic, and antiferromagnetic. After the geometry optimizations, the most stable 2D FeS$_2$ oriented in [100] and [110] directions are the antiferromagnetic one, being ~100 and 63 meV/atom lower than the ferromagnetic one, and ~240 and 285 meV/atom lower than the diamagnetic one, respectively. However, 2D FeS$_2$ oriented in [111] direction has a stable ferromagnetic state, being 9 and 17 meV/atom lower than the antiferromagnetic and diamagnetic states respectively. Similarly, the magnetic state for a bilayer of FeS$_2$ oriented in [111] direction (24 atoms, 8 FeS$_2$ formulas) is also calculated, and we found that the ferromagnetic state is still the most stable, being 7 and 81 meV/atom lower than the antiferromagnetic and diamagnetic states respectively (see **Figure S6** for spin density of states). The Curie temperature (Tc) and the exchange interaction parameter (J) for both a monolayer and the bilayer FeS$_2$ oriented in [111] are estimated based on the Heisenberg hamiltonian and the mean field approximation (see Supplementary Information for the detailed calculations). The estimated Curie temperature (Tc) and the exchange interaction parameter (J) for the monolayer (bilayer) FeS$_2$ are, respectively, 110.05 K (83.86 K) and 2.71 meV (1.24 meV).

In **Figure 4a**, the geometry optimized unit cell of the most stable 2D FeS$_2$ oriented in [100], [111], and [110] directions are presented, i.e., in their antiferromagnetic, ferromagnetic, and antiferromagnetic states, respectively. The respective lattice parameters (**a** and **b**) obtained for these structures are 5.58 x 5.58 Å ($\alpha=\beta=\gamma=90°$, point/space group C2v/Pba2), 6.40 x 6.70 Å ($\alpha=\beta=90°$, $\gamma=122°$, point/space group C$_1$/P1), and 5.35 x 8.10 Å ($\alpha=\beta=90°$, $\gamma=94°$, point/space group C$_1$/P1) for 2D FeS$_2$ oriented in [100], [111], and [110] directions, respectively. As for the thickness, from our calculations, it is found out that the 2D FeS$_2$ sheets oriented in [111] and [110] directions are having a thickness of 3.5 and 3.8 Å respectively, whereas [100] oriented is single atom thick (see **Figure 4a**). Regarding the bond lengths, we have obtained the values around 2.14-2.18 Å for Fe-S and ~2.19 Å for S-S, which are shorter than the same reported for bulk FeS$_2$.[43]

In **Figures 4b** and **4c** we present, the spin density (different colors indicated the spin up and down configurations) and the density of states (DOS) of the 2D FeS$_2$ oriented in [100], [111], and [110] directions in their antiferromagnetic, ferromagnetic, and antiferromagnetic states, respectively. The spin density calculations show that the major contribution to the magnetic response of each monolayer system comes from the Fe atoms with magnetic moments varying between 1.1 to 2.0 $\mu_B$, (see Supplementary Information **Table S4-12**) and 12.61$\mu_B$ for [111] oriented bilayer system (**Table S13**). As for the DOS, it is interesting to notice that both ferromagnetic as well as antiferromagnetic states indicate a metallic behavior. In **Figure 4d** we present a schematic view of the net magnetization per unit cell for 2D FeS$_2$ oriented in [100] and [110] directions. To sum up, the DFT calculations indicate that the 2D FeS$_2$ oriented in [111] direction is the easier to exfoliate and its ferromagnetic state is energetically stable compared to the antiferromagnetic and diamagnetic ones. However, more investigations are essential to unravel the origin of ferromagnetism in 2D $FeS_2$ and the type of exchange interaction responsible for the long-range magnetic ordering.

The theoretical findings are evaluated experimentally by low temperature hysteresis measurements in a superconducting quantum interference device (SQUID) magnetometer. The supernatant of the solution consisting of ultrathin $FeS_2$ sheets obtained after ultracentrifugation was vacuum filtered and dried under vacuum at room temperature and a few milligrams of the obtained sample was subjected to SQUID magnetization measurements. Interestingly, exfoliated pyrite has an anomalous ferromagnetic like magnetic ordering as one could clearly see from the characteristic low temperature M-H hysteresis at 1.8 K (**Figure 5a**) with large coercive field around 700 Oe (**Figure 5a inset**). However, the M-H hysteresis is not saturating even at the maximum applied field of 7T. This could be due to the presence of antiferromagnetic (100) and (110) faceted sheets which cannot be ruled out. Such a behavior is well in accordance with our theoretical calculations. However, absence of a clear magnetic transition and non-diverging magnetization versus temperature FC-ZFC measurements (Supplementary Information **Figure S7a**) rule out the possibility that the observed anomalous magnetism is solely due to ultra-thin exfoliated sheets of pyrite. Contributions from the uncompensated spins due to oxide (such as $Fe_2O_3$ or $Fe_3O_4$) formation on the surface of the exfoliated pyrite sheets could also be accountable for the exfoliation induced magnetism. A weak Fe2p3/2 XPS mode centered on 709 eV (**Figure 2c**) corresponding to surface states also points to the inevitable surface oxide formation during exfoliation. At room temperature, it becomes paramagnetic as evident from **Figure 5b** having a linear M versus H behavior (**Figure 5b inset**). This transition from a ferromagnetic behavior at low temperature to a paramagnetic one at room-temperature is substantiated by theoretical Curie temperature calculations.

Extended magnetic measurements showcasing the diamagnetic response of the bulk precursor and magnetization versus temperature are provided in the supplementary information (**Figure S7b**). Even if the SQUID magnetic measurements undoubtedly confirms the exfoliation induced anomalous magnetic ordering in exfoliated pyrite sheets, experimental determination of transition temperature and ratification magnetic ground states for various theoretically predicted orientations demand monolayer sensitive techniques such as MOKE, MCD or Nitrogen Vacancy Center Magnetometry and could be even useful for studying thickness dependent changes in magnetic ordering.[44]

**Conclusions**

In short, we have successfully isolated ultra-thin n-vdW $FeS_2$ nanosheets from natural iron sulfide (pyrite $FeS_2$) by means of liquid-phase exfoliation. DFT calculations predicted that a (111) faceted three atoms thick $FeS_2$ sheet is stable and can be exfoliated quite easily by virtue of its least energy of exfoliation. It consists of a layer of iron atoms sandwiched between two layers of sulfur atoms. Ultra-thin $FeS_2$ nanosheets are experimentally found to be retaining the crystallinity of the parent pyrite structure. Spin polarized density functional theory calculations predicted a ferromagnetic ground state for (111) faceted 2D $FeS_2$ and an anomalous ferromagnetic like behavior of exfoliated pyrite sheets is observed experimentally. The observed magnetism of exfoliated pyrite is attributed to both the contribution of the exfoliated sheets as well as the uncompensated spins of the surface oxides. The exfoliation induced anomalous magnetic ordering in ultra-thin n-vdW crystals could have excellent implications in both fundamental magnetism as well as in future spintronic applications.

**Supporting Information**

Optical images of pristine and exfoliated pyrite sheets, XPS Survey Spectra, Extended AFM, TEM, Magnetization, and theoretical analysis and EMP data and analysis.

**Acknowledgments**

This material was based upon work supported by the Air Force Office of Scientific Research under award number FA9550-18-1-0072 (P.M.A., A.B.P.). A.P.B., V.S. and R.R.N. would like to thank European Research Council (Contract 679689) for the support. EFO and DSG would like to thank the Brazilian agencies CNPq and FAPESP (Grants 2016/18499-0, and 2019/07157-9) for financial support. Computational support from the Center for Computational Engineering and Sciences at Unicamp through the FAPESP/CEPID Grant No. 2013/08293-7 and the Center for Scientific Computing (NCC/GridUNESP) of São Paulo State University (UNESP) is also acknowledged. L.M.S acknowledges CAPES (Coordination for the Improvement of Higher Education Personnel) under the Brazilian Ministry of Education for the financial support.

**Author Contributions**

A.B.P., A.P.B., E.F.O., and P.M.A envisioned and executed the experiments. E.F.O. and D.S.G. did the theoretical analysis and computations. G.G., F.C.R.H., and A.B.P. did the electron microscopy, analysis, description and commented in the manuscript. G.C. provided the natural ore sample and analyzed (EPMA) the parent crystal. V.S. and L.M. S. performed AFM imaging. T. P. and N.C. helped in preparing the samples and joined for discussions and interpreting the results. P.M.A., R.R.N., D.S.G. and R.V. guided the research. All the authors contributed towards writing the manuscript.

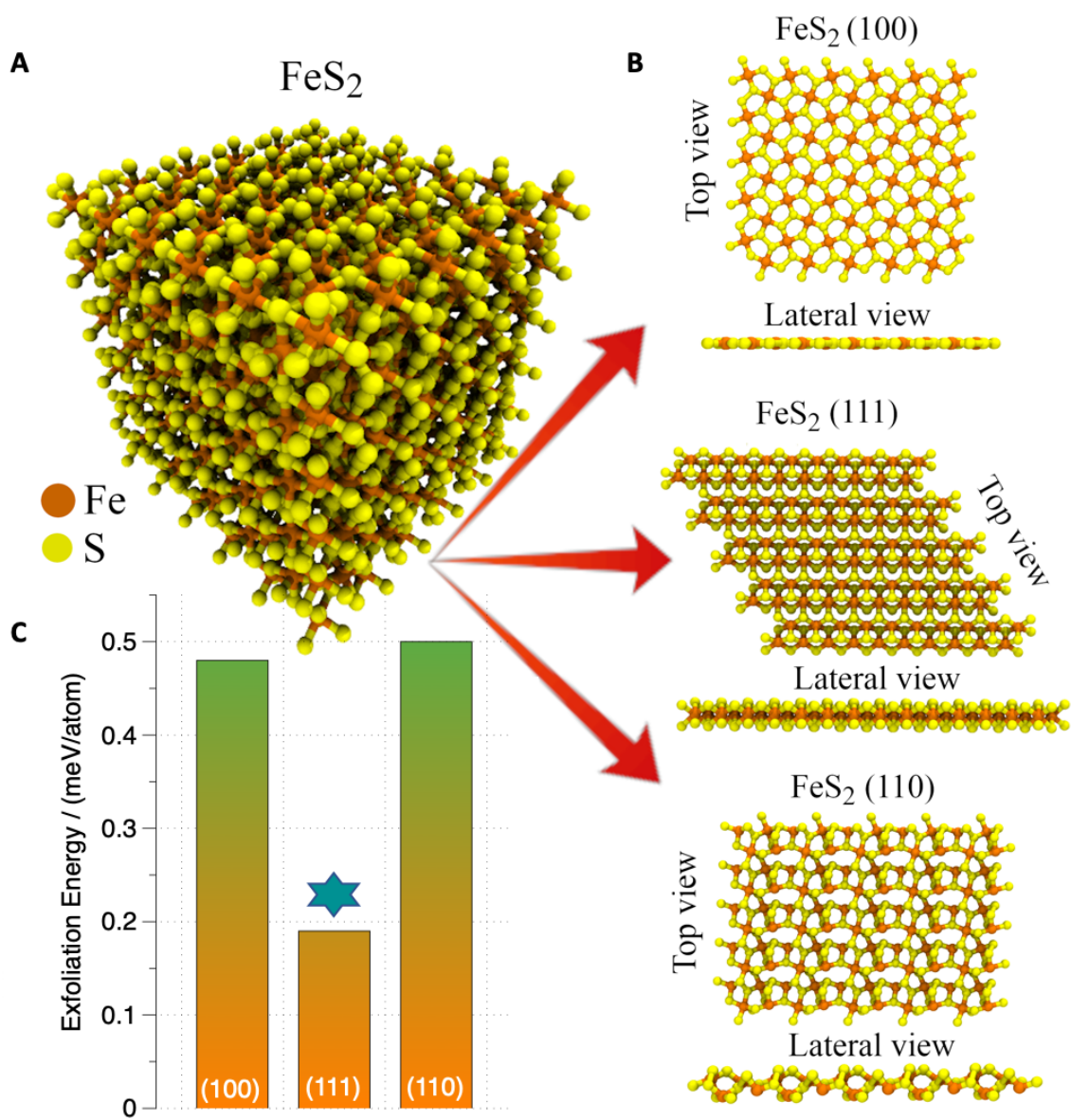

**Figure 1**. Feasibility of liquid exfoliation of Iron pyrite (FeS$_2$). **(a)** Bulk iron pyrite crystal **(b)** Based on the cleavage and parting in the bulk crystal, three planes are likely to be exfoliated and are (100), (111) and (110). **(c)** The energy required for exfoliation is calculated, and (111) plane is found to show the least energy and hence is expected to be abundant in the exfoliated samples.

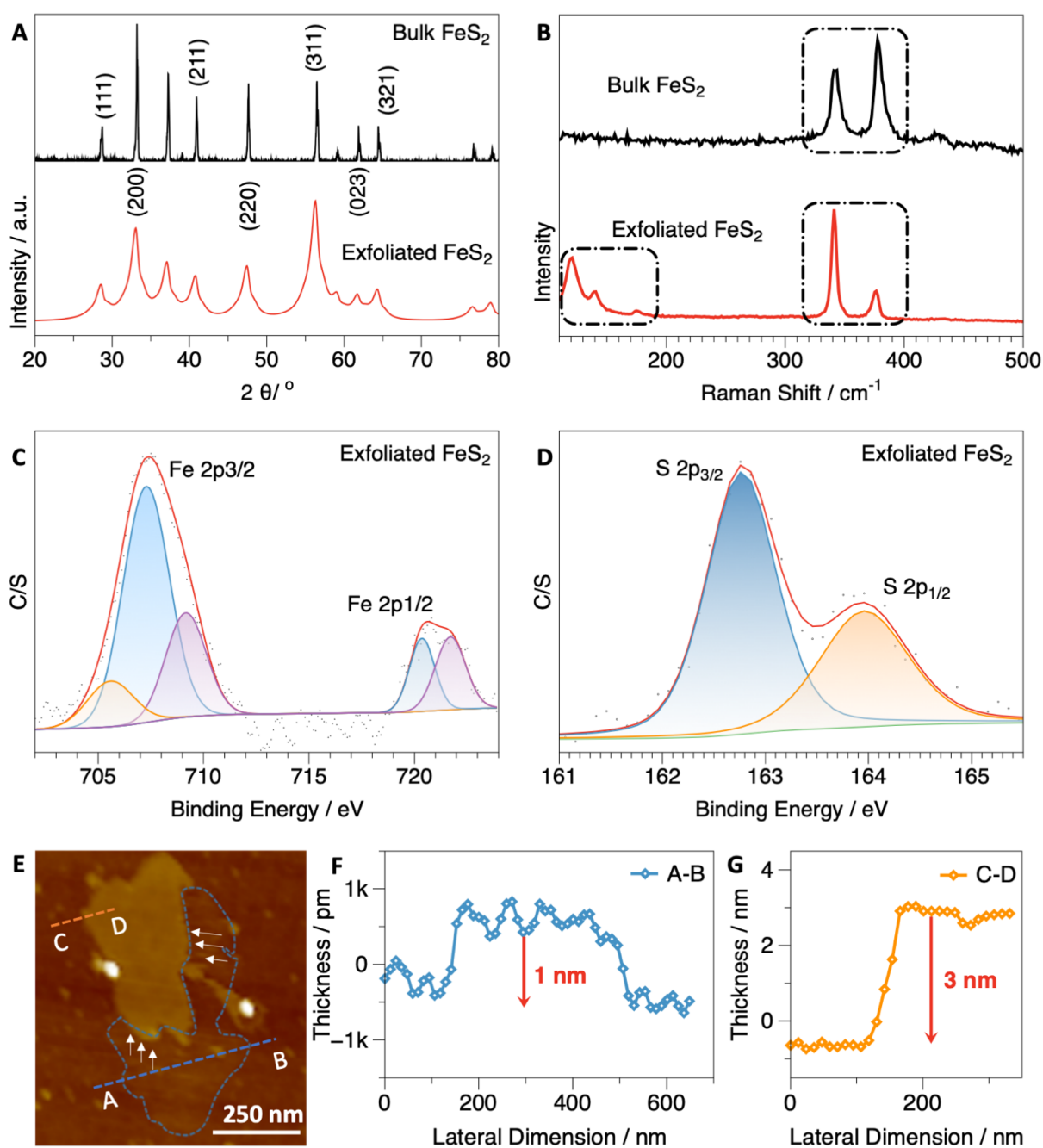

**Figure 2.** Structural and elemental analysis: **(a)** X-ray diffraction spectrum of bulk (top) and exfoliated $FeS_2$ (bottom) - the diffraction peaks are less intense and broad for the exfoliated samples compared to the bulk counterpart. Also, the intensity ratio of the diffraction peaks is different due to the confinement effects. **(b)** Raman spectra of bulk (top) and exfoliated $FeS_2$ (bottom). The intensities of prominent Raman modes are varied from bulk to exfoliated sheets and the lower frequency Raman modes present in exfoliated sheets are characteristic of vdW like layers. **(c)** High resolution XPS spectra corresponding to Fe2p and **(d)** S2p of exfoliated $FeS_2$. **(e)** AFM topography image and corresponding line scan curves showing the thickness of **(f)** 1 nm and **(g)** 3 nm.

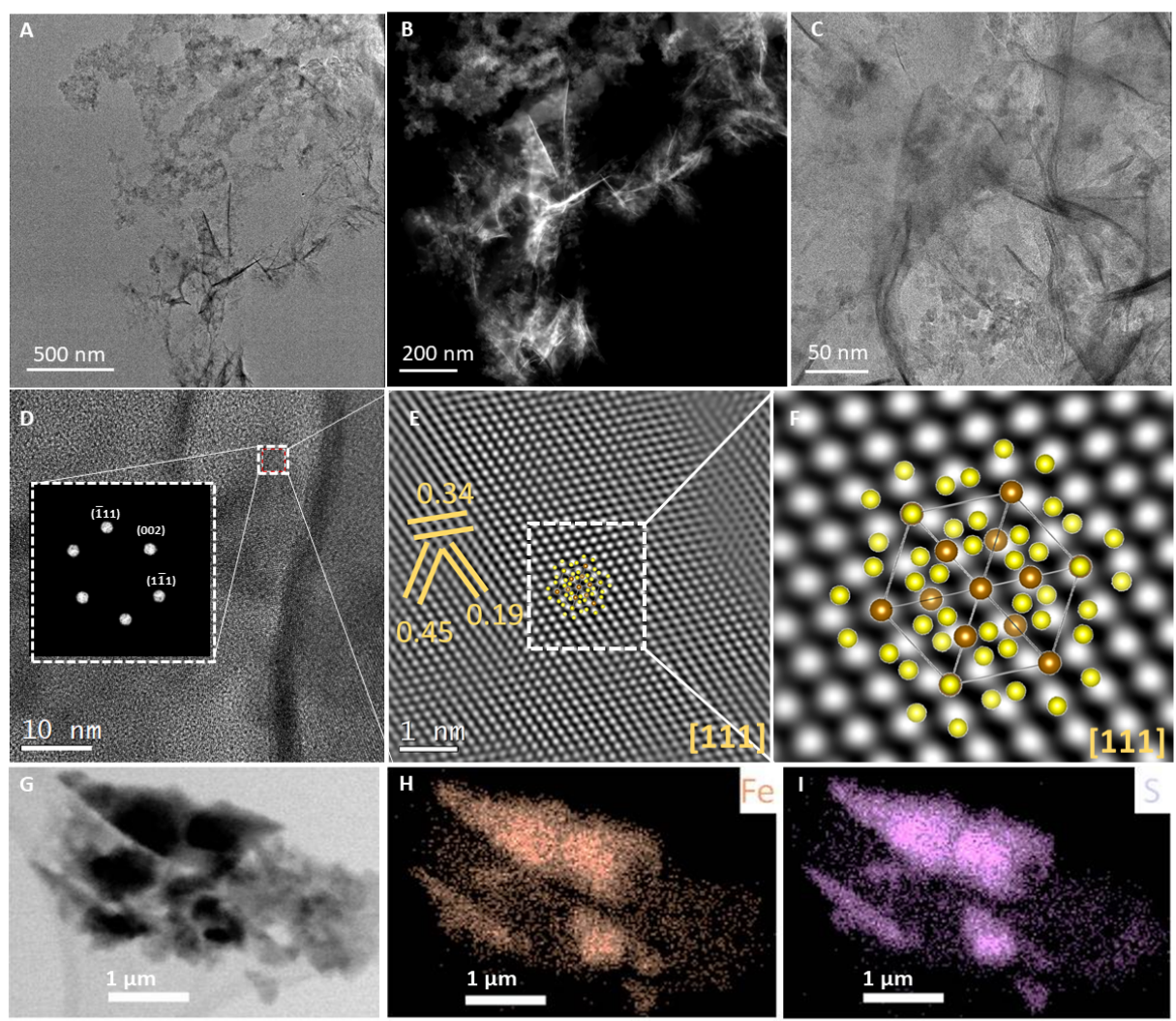

**Figure 3.** Transmission electron microscopy **(a)** Low magnification showing the overall area of the flakes **(b)** moderate and **(c, d)** high magnification micrographs of exfoliated FeS2 samples. The corresponding FFT patterns (d inset) confirms [111] preferential orientation of the exfoliated sheets corresponding to (111) planes. **(e)** high- resolution micrograph with atomic resolution **(f)** FeS$_2$ projection modeled in Vesta® software and superimposed over a zoom-in portion from **(e)** demonstrating match between simulated and IFFT structures. **(g)** The backscattering image of an ultra-thin sheet and (h) the elemental mapping of iron (Fe) and (i) of sulfur (S).

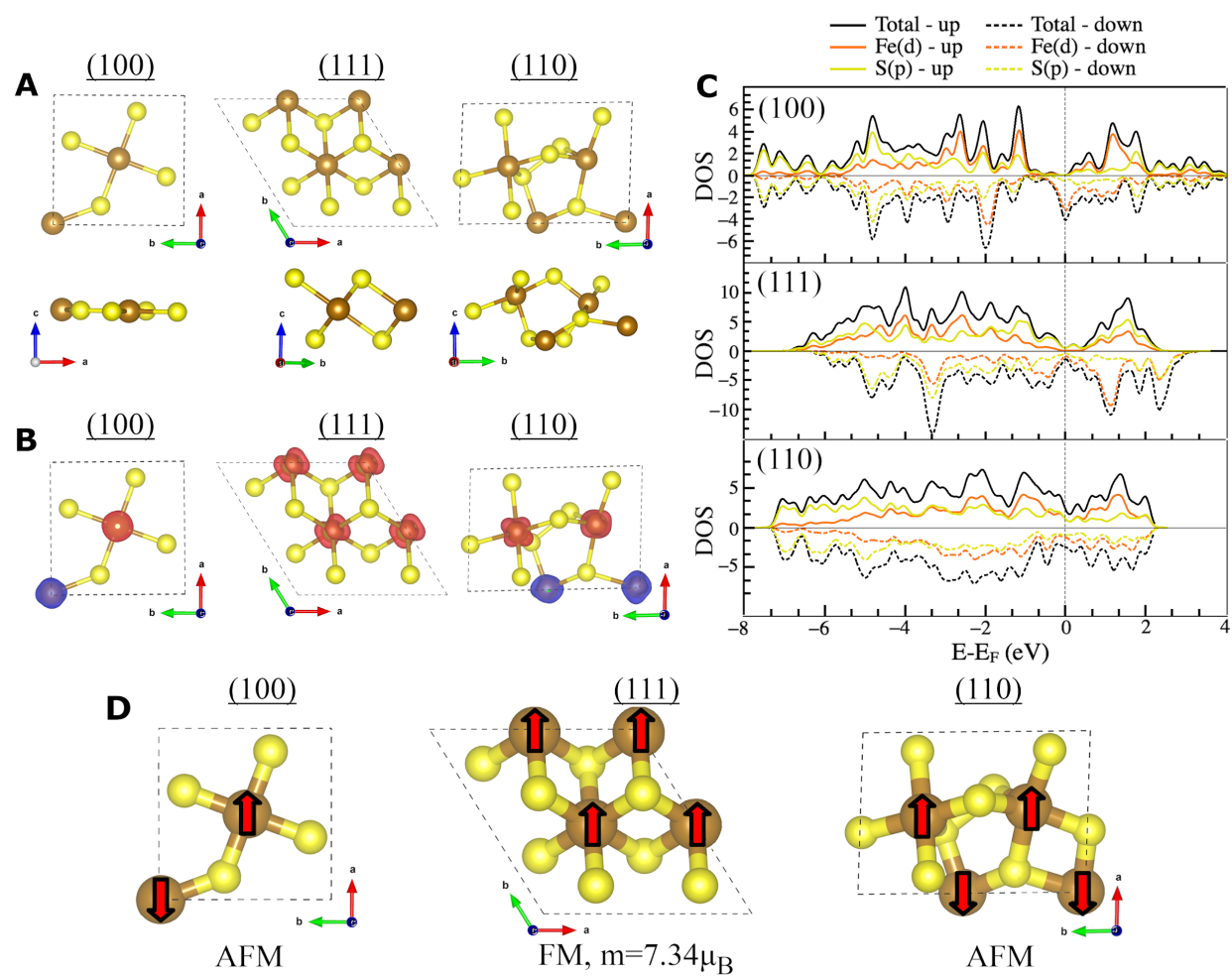

**Figure 4**. DFT modelled structures, the density of states (DOS) and spin density and magnetization per unit cell: **(a)** Optimized geometry in a top and side view, **(b)** spin density in a top of the view, **(c)** the total density of states (DOS) from -8.0 to 4.0 eV, and **(d)** schematic of net magnetization per unit cell of 2D $FeS_2$ oriented in [100], [111], [110] directions in their antiferromagnetic, ferromagnetic, and antiferromagnetic states, respectively.

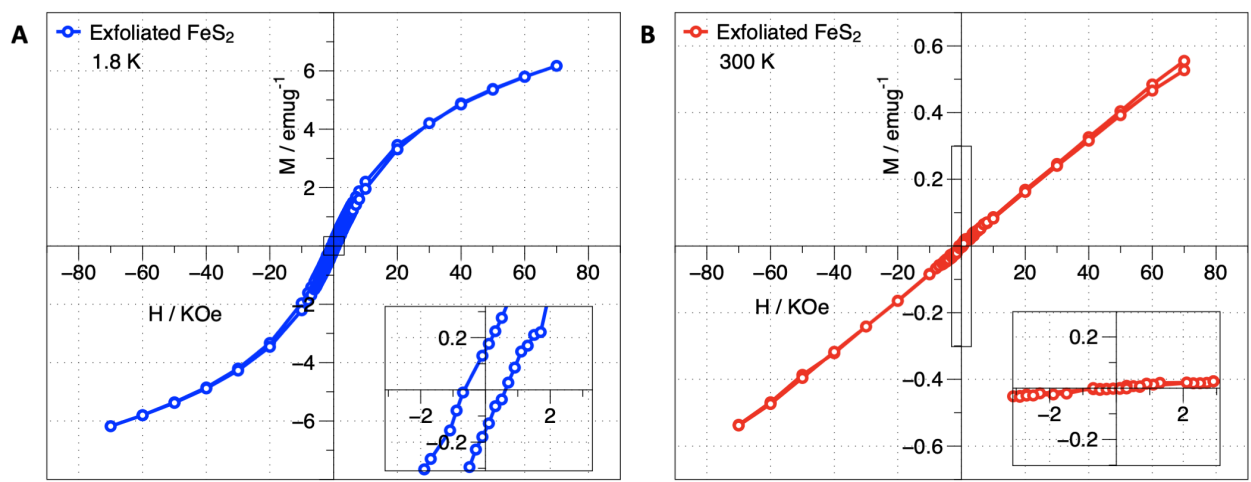

**Figure 5.** Low temperature, 1.8K and room temperature, 300K M-H hysteresis measurements (**a** & **b** respectively) of exfoliated $FeS_2$ with the corresponding enlarged views in the inset.